# Nonlinear Negative Refraction by Difference Frequency Generation

JIANJUN CAO[1], DONGYI SHEN[1], YAMING FENG[1] AND WENJIE WAN[1, 2*]

[1]*MOE Key Laboratory for Laser Plasmas and Collaborative Innovation Center of IFSA, Department of Physics and Astronomy, Shanghai Jiao Tong University, Shanghai 200240, China*
[2]*The State Key Laboratory of Advanced Optical Communication Systems and Networks, University of Michigan-Shanghai Jiao Tong University Joint Institute, Shanghai Jiao Tong University, Shanghai 200240, China*
*\*Corresponding author: wenjie.wan@sjtu.edu.cn*

**Negative refraction has attracted much interest for its promising capability in imaging applications. Such an effect can be implemented by negative index meta-materials, however, which are usually accompanied by high loss and demanding fabrication processes. Recently, alternative nonlinear approaches like phase conjugation and four wave mixing have shown advantages of low-loss and easy-to-implement, but associated problems like narrow accepting angles can still halt their practical applications. Here we demonstrate theoretically and experimentally a new scheme to realize negative refraction by nonlinear difference frequency generation with wide tunability, where a thin BBO slice serves as a negative refraction layer bending the input signal beam to the idler beam at a negative angle. Furthermore, we realize optical focusing effect using such nonlinear negative refraction, which may enable many potential applications in imaging science.   © 2015 Optical Society of America**

Negative refraction (NR) bending light in a reversed manner as opposed to the normal refraction has continuously attracted growing interests from many fields including optics, acoustics and microwaves [1-8], for its promising applications in imaging, cloaking, sensing [9-14]. Such a phenomenon has been realized in many formats including photonic crystals [8], metal thin films [10], meta-materials [3-7, 11], while high losses especially in optical regime from metallic materials: the key elements enabling NR, and the demanding technologies to fabricate these nano/micro structures, strongly limit NR's practical applications. Alternatively, an effective NR has been proposed in nonlinear optics by using nonlinear optical processes such as phase conjugation, time reversal and four wave mixing (4WM) in order to achieve effective NR between incident waves and nonlinear generated ones. For example, NR has been demonstrated with signal beams and 4WM beams in a 4WM scheme with thin 3rd order nonlinear slab such as metal, graphite thin film and glass [15-19]. Furthermore, one revolutionary imaging technique with potentials to achieve super-resolution microscopy has been realized by exploring the phase matching conditions during 4WMs in Ref. [20-21]. However, in these nonlinear NRs, some minor problems such as narrow phase matching angle may still hold off their applications in imaging.

In this letter, we propose and experimentally demonstrate a new way to achieve NR based on nonlinear difference frequency generation (DFG). With a thin slice of BBO crystal containing second-order nonlinearity, an infrared signal beam interacts with the pump beam nonlinearly through DFG process to give rise to the visible idler beam which is negatively refracted with respect to the signal to fulfill the phase matching. These negatively refracted idler beams can focus and form the image of the object illuminated by the signal beams. We analyze the focusing behavior in both non-collinear and collinear configurations where one-dimensional focusing and two-dimensional focusing can be obtained respectively. The phase matching condition governs the negatively refracted angle as a function of the incident angle, which is tunable by tilting the nonlinear material. This gives us a degree of freedom to select suitable spatial frequencies for imaging purposes. Our method is promising to be employed in nonlinear converted infrared microscopy and angle-resolved spectroscopy.

DFG is a second order nonlinear process where the strong pump beam at frequency $\omega_1$ interacts with the signal beam at frequency $\omega_2$ to give rise to the idler beam at frequency $\omega_3 = \omega_1 - \omega_2$ according to the energy conservation law. This process can take place efficiently when the phase matching condition ($\vec{k_p} = \vec{k_s} + \vec{k_i}$) is satisfied, where $\vec{k_p}, \vec{k_s}$ and $\vec{k_i}$ are the corresponding wave vectors of the pump, signal and idler respectively. In imaging science, DFG can be applied in converting infrared waves into visible ones [22-25] over a wide spectrum as long as the signal and the idler beams fulfill the energy conservation law; on the other hand, the phase matching condition of DFG enables applications in the phase conjugation [26-28]. Moreover, recent works in degenerated 4WM based NR flat lens have opened up a new route towards super-resolution imaging [20-21], and degenerated 4WM process in the 3rd order nonlinear medium is nearly identical to DFG in the 2nd order nonlinearity if the two pump photons in 4WM are replaced by one photon in DFG. Hence, we would like to explore DFG for negative refraction and its potential applications in imaging by considering DFG's flexible phase matching schemes and the higher conversion efficiency.

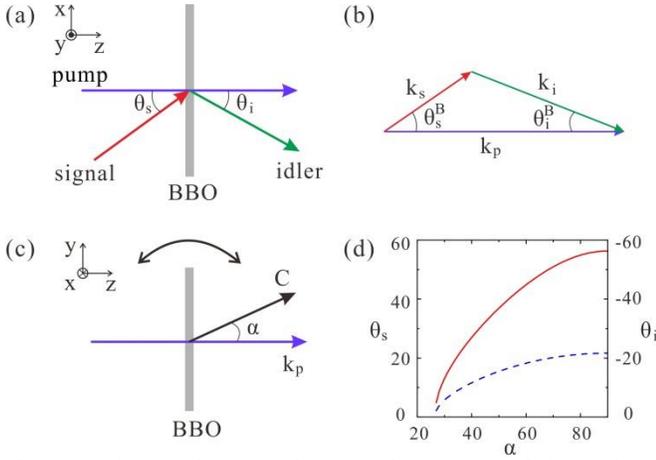

Fig. 1. (a) Schematic of negative refraction based on DFG. The idler beam is negatively refracted relative to the incident signal beam by nonlinearly interacting with the incident pump beam in the BBO slice. The incident angle between the pump and the signal is defined as $\theta_s$ while the refracted angle between the pump and the idler is defined as $\theta_i$. (b) The phase matching condition of the DFG process. The wave vectors need to compose a triangle to satisfy the phase matching condition $\vec{k_p} = \vec{k_s} + \vec{k_i}$, which leads to the negative relationship between $\theta_i^B$ and $\theta_s^B$ which are angles inside the BBO slice. (c) Schematic of the tunable angle α between the propagation direction of the pump beam and the principle C –axis of the BBO crystal. (d) The incident angle $\theta_s$ (solid red line) and the refracted angle $\theta_i$ (dash blue line) as a function of α.

DFG based nonlinear negative refraction scheme is shown in Fig. 1(a): the pump beam at frequency $\omega_1$ is incident normally onto a thin Beta barium borate (BBO) slice with the second order nonlinear susceptibility $\chi^{(2)}$, meanwhile, a secondary signal beam at frequency $\omega_2$ is launched obliquely at angle $\theta_s$ onto the same spot, this gives rise to the idler beam at frequency $\omega_3$, which can be negatively refracted at angle $\theta_i$ with respect to the signal's incident angle in order to fulfill the phase matching condition. Here, the type I (e → o + o) [29] phase matching condition determines the relationship between the angles as:

$$k_p = k_s \cos\theta_s^B + k_i \cos\theta_i^B, \quad (1)$$

$$k_s \sin\theta_s^B = -k_i \sin\theta_i^B, \quad (2)$$

where $k_p = 2\pi n_p^e/\lambda_p$, $k_s = 2\pi n_s^o/\lambda_s$ and $k_i = 2\pi n_i^o/\lambda_i$ (n indicates the refractive index and λ is the wavelength). $\theta_s^B$ and $\theta_i^B$ are angles inside the BBO slice. These two angles are related to the angles in air by Snell's law:

$$\sin\theta_s = n_s^o \sin\theta_s^B, \quad (3)$$

$$\sin\theta_i = n_i^o \sin\theta_i^B. \quad (4)$$

Insert Eqs. (3) and (4) into Eq. (2), a nonlinear refraction law is obtained:

$$\frac{\sin\theta_s}{\sin\theta_i} = -\frac{\lambda_s}{\lambda_i}, \quad (5)$$

The negative nature of $\theta_i$ relative to $\theta_s$ is clearly shown in Eq. (5). Interestingly, unlike the negative refraction effect caused by four-wave mixing [20] where nonlinear negative refraction is restrict to a single scheme defined by the phase matching condition, here the negative refraction is tunable thanks to the birefringent nature of BBO crystal [29], where $n_p^e$ can be adjustable by varying the incident angle α between the pump and the principal C-axis of the BBO crystal as shown in Fig. 2(c), this modifies the phase matching conditions stated in Eq. (1) and Eq. (2). As a result, both the signal angle $\theta_s$ and the idler angle $\theta_i$ in the phase matching triangle can be varied as a function of α shown in Fig. 2(d), which gives an extra tunability in contrast to the 4WM case [20]. Physically, such process can be achieved by titling the BBO crystal in the y-z plane (Fig. 1c), which only modifies $n_p^e$, meanwhile, both the pump and the signal still lay in the x-z plane (Fig. 1a), the generated idler beam, also in the x-z plane, will be affected through the modified phase matching. For an imaging system, this simple titling mechanism may enable tunable numerical aperture (NA), enhancing the imaging resolution [30-31].

Experimentally, we verify this NR phenomenon by the setup illustrated in Fig. 2(a): a Ti:Sapphire femtosecond laser delivering pulses of duration ~75 fs and central wavelength 800 nm is injected into an optical parametric amplifier which outputs pulses of similar duration of wavelength 1300 nm. The un-depleted pulses at 800 nm are up-converted into pulses at 400 nm by a BBO slice cut at 22.9° for Type I phase matching through second harmonic generation process. After passing through a 600 nm short pass filter, the pulses at wavelength $\lambda_p = 400$ nm are served as the pump beam in the DFG process and the pulses at wavelength $\lambda_s = 1300$ nm are employed as the signal beam. A delay line is added to ensure the pump and the signal's synchronization in time. In the experiments, the signal and the pump are incident into a 200μm-thick BBO (cut at 44.3°, marked as BBO2 in Fig. 2(a)) with an intersection angle of $\theta_s$, then the BBO slice is titled along the vertical transverse plane till the phase matching condition is satisfied, which enables the negative refraction of the idler beam: as show in Fig. 2(b), the screen of the output beams collects the blue spot of the pump, the orange spot indicating the idler, while the left red spot is the signal's however it mixes both the signal's at the fundamental wavelength 1300 nm and its second harmonic at wavelength 650 nm. The DFG conversion efficiency is measured about $2.3 \times 10^{-3}$. Note that, the idler is on the opposite side to the signal indicating the occurrence of NR. One advantage of DFG based NR over the 4WM case is the wide tunable angles: here for a given $\theta_s$, by titling the BBO slice, we can always find a suitable phase matching condition (basically satisfying Eq. (1) ) enabling DFG. The measured relationship between $\sin \theta_i$ and $\sin \theta_s$ is shown in Fig. 2(c), which fits well with the theoretical calculations by Eq. (5). The signal's incident angle can vary from 0 degree – 43 degree in the current experiment. Furthermore, using sub-wavelength-thick material which gives more freedom in phase matching, it is possible to convert signal waves with high spatial wave vectors into idler waves with small ones by DFG, allowing super resolution imaging in analogy with the methods by four wave mixing as discussed in ref. [30-32].

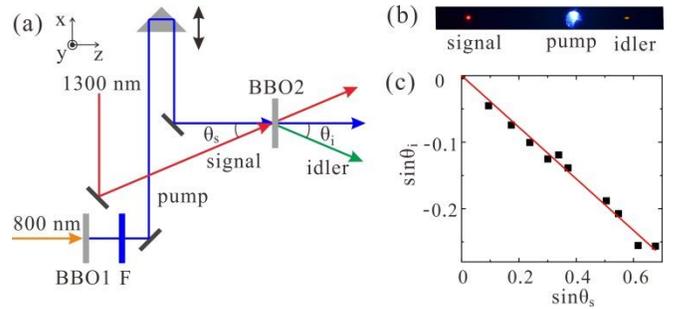

Fig. 2. (a) Experimental setup to verify the negative refraction phenomenon by DFG. "BBO1" is a BBO slice cut at 22.9° to convert the pulses at 800 nm into pulses at 400 nm. " F " is a 600 nm short pass filter. "BBO2" is a BBO slice cut at 44.3° where the DFG process takes place. (b) The imaged spots of the signal, pump and idler on a screen, respectively. The idler is on the opposite side of the pump relative to the signal, confirming the negative refraction behavior. (c) The relationship between $\sin\theta_i$ and $\sin\theta_s$. The squares are experimentally measured data and the solid red line is the theoretical fitting line calculated by Eq. (5).

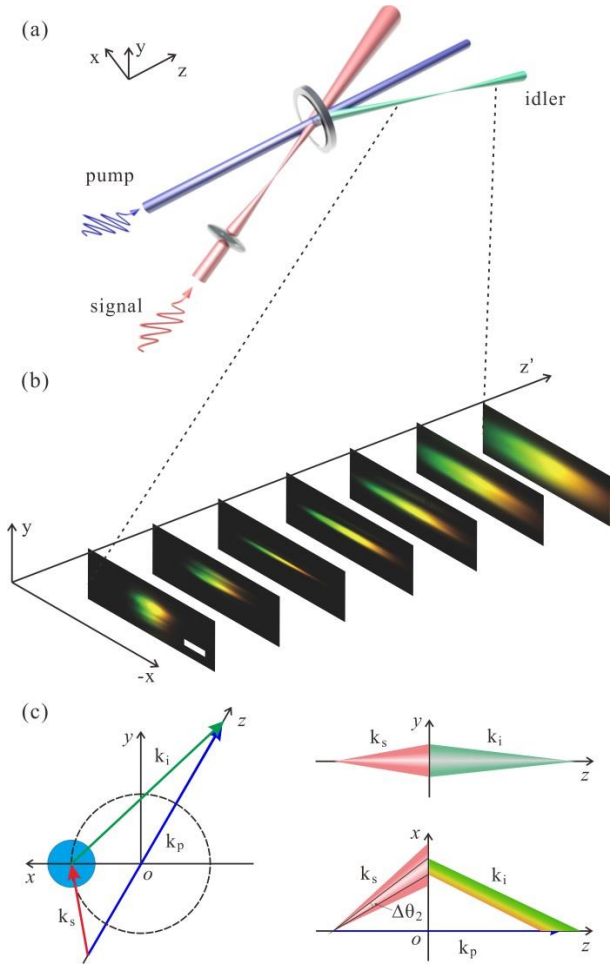

Fig. 3. Imaging a point using the negative refraction effect based on DFG in a non-collinear setup. (a) Schematic of the non-collinear experimental setup. (b) The captured images of the idler along its travelling path from 7.5 cm to 27.8 cm. z' is the direction of the idler beam. (c) Diagram of phase matching condition composes a ring in three dimension vector space, where the ring segment in the y-z plane allows a better focusing while the small beam divergence combined with the frequency spreading of the pluses gives rise to multicolor DFG generation in the x-z plane.

DFG based NR potentially can be used in imaging similar to the 4WM case [20-21], where a thin flat lens can be realized for imaging by focusing multiple beams along the phase matching cone of 4WM. Similar effect can be expected with DFG process for a better conversion efficiency and tunability as shown above. But prior to realization of imaging, we first demonstrate that the BBO slice can serve as a flat lens to focus a diverging beam, where the focusing function can be explored further for imaging purposes. Here in a non-collinear configuration shown in Fig. 3(a), the signal beam with wavelength of 1300nm is incident on the BBO slice at angle $\theta_s = 17.2°$ in the x-z plane while the pump beam maintains at the normal incidence. In order to image a single spot, we create a signal spot 6 cm in front of BBO slice by using a lens with focal length of 6 cm. BBO is proper tilted in the y-z plane to ensure the efficient DFG process, so that idler beam can be generated and its spatial geometry is captured by a CCD camera. The captured images of the idler along its travelling path from 7.5 cm to 27.8 cm are shown in Fig. 3(b), where the idler beam is first focused into a single stripe around 12.9 cm and then diverging after this point in y-axis, meanwhile the generated DFG spot keeps spreading out along x-axis without any focusing. Such phenomenon results from a better phase matching scheme on the vertical plane (y-z plane), giving rise to a better focusing.

Here, the DFG process fulfills the type I (e → o + o) phase matching condition [29]. The angle between the pump beam and the principle axis-$C$ of the BBO crystal is fixed, giving the refractive index of $n_p^e$ for the pump. While the refractive indices of the idler and signal maintain at $n_i^o$ and $n_s^o$ due to the o-polarization nature. Correspondingly, the absolute values of $k_p, k_i, k_s$ during the phase matching in the BBO crystal are also determined so that the endpoints of $k_s$ form a ring in three-dimensional vector space as shown in Fig. 3(c). In the previous works of 4WM, such phase matching ring was explored for focusing and imaging purposes with any incident waves fulfilling such phase matching conditions [20]. Here the non-isotropic phase matching ring segment allows a better phase matching in the y-z plane rather than the x-z plane in Fig. 3(c), this enables a better focusing from multiple angled DFG waves in the y-z plane than those in the x-z plane. Moreover, the small angle beam divergence combined with the frequency spreading of the pluses gives rise to multicolor DFG generation in the x-z plane. Similar phenomenon has been fully discussed in the 4WM experiments [20-21].

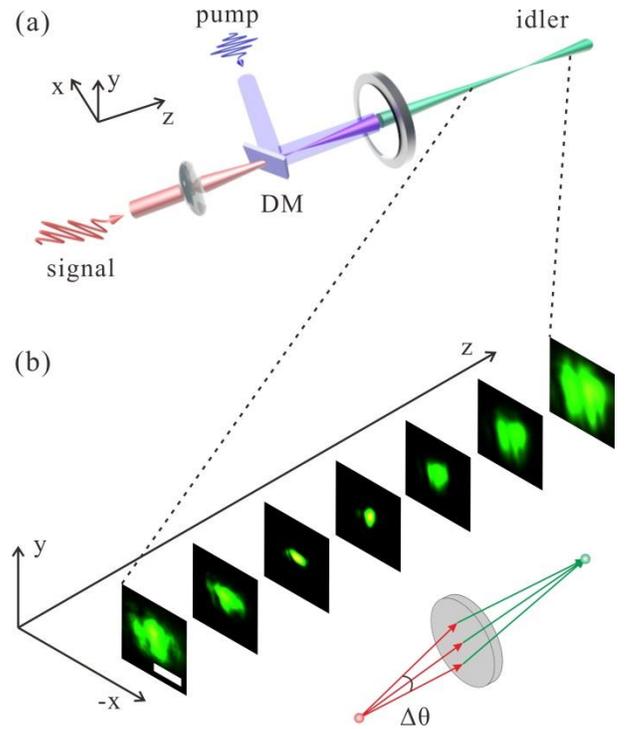

Fig. 4. Imaging a point using the negative refraction effect based on DFG in a collinear setup. (a) Schematic of the collinear experimental

setup. DM is a 900 nm long pass dichroic beam splitter. (b) The captured images of the idler along its travelling path from 7.5 cm to 27.8 cm. The insert shows the cross section of the focusing process with a small angle tolerance of Δθ during the collinear phase matching scheme.

To realize two dimensional focusing in both transverse axes, we construct a collinear setup as shown in Fig. 4: now the signal and the idler are incident onto the BBO slice collinearly. This can be done also by titling the BBO slice till reaching the collinear phase matching for both incoming beams. The imaging spot along the path of the signal was still 6 cm in front of the BBO slice. The idler beam is captured by a CCD camera after filtering the pump and the signal beams using a 600 nm short pass filter and a 425 nm long pass filter. Images of the idler from 7.7 cm to 25.7 cm are shown in Fig. 4(b), where the idler focuses around the distance of 13.7cm both in x-axis and y-axis. Unlike the above non-collinear case where the phase matching only allows focusing on the vertical plane, here the collinear configuration allows focusing in both axes. Partially, this is because the phase matching ring in the non-collinear case (Fig. 3c) has reduced to a dot in this collinear configuration, which is isotropic on the transverse plane. Meanwhile, the phase matched DFG process here allows a small angle tolerance Δθ, which is measured to be ~1° experimentally. Such angle tolerance of the phase matching angle immediately enables multiple beams inside a small angle spreading cone (Fig. 4b) to focus their DFG waves on the other side of the BBO crystal. Both in the non-collinear and collinear setups, the signal beams diverging from the focus spot undergo a linear optical path. However, the idler beams generated by the DFG are focused to a point, undergoing a nonlinear optical path, which are negatively refracted relative to the signals. The BBO slice acts as a negative refraction boundary condition bending the optical path of the signal into the idler. Hence, such nonlinear NR effect can be further explored for imaging applications in the future.

Here, the DFG process is performed in a slab of finite thickness, where full phase matching condition should be satisfied, which means that Eqs. (1) and (2) should be satisfied simultaneously. This restricts the signal's angle to a certain angle. But imaging can still be performed in the non-collinear setup using part of the phase matching ring as shown in Fig. 3 and in the collinear setup using the angle tolerance Δθ as shown in Fig. 4. And if the thickness of slab reduced to subwavelength, due to the subwavelength propagation length, phase matching does not have a chance to make an impact on the efficiency of the generation of a homogeneous wave, only Eq. (2) needs to be satisfied. This absence of the phase matching requirement allows conversion of a wide range of k-vectors of the signal wave at $\omega_2$ to a set of k-vectors of a generated idler wave at $\omega_3$ [33-34]. This all-angle negative refraction may have wider applications if the problem of low conversion efficiency can be overcome.

At last, we would like to comment on the merits of DFG based nonlinear NR compared with the 4WM case. In an imaging process, light scattered by an object can form images by the aid of a lens by refracting those light having spatial frequencies smaller than the numerical aperture. In a nonlinear flat lens based on 4WM, only a cone of beams with particular incident angles can be negatively refracted and form images, which is not tunable due to the rigid phase matching condition. However, with DFG case, such incident angles are adjustable as shown above by titling the BBO slice to ensure the phase matching conditions at various incidences. Such flexibility of DFG may eventually be useful for some imaging applications which require fine angle-resolved capability in spectroscopy [35].

In conclusion, we demonstrate experimentally nonlinear negative refraction by difference frequency generation for the first time. We have shown the tunability of DFG through tuning the phase matching conditions. Both 1D and 2D focusing have been realized in the non-collinear and collinear configurations. We expect that this method can be further explored for the applications in infrared microscopy and angle-resolved spectroscopy.

**Funding.** National Natural Science Foundation of China (Grant No. 11304201, No. 61475100), the National 1000-plan Program (Youth), Shanghai Pujiang Talent Program (Grant No. 12PJ1404700), Shanghai Scientific Innovation Program (Grant No. 14JC1402900). Shanghai Scientific Innovation Program for International Collaboration (Grant No. 15220721400).

**Acknowledgment**. We are grateful to Prof. Xianfeng Chen for using ultrafast laser facility.